\documentclass[prc,aps,a4paper,groupedaddress,superscriptaddress,nofootinbib,showpacs
,preprintnumbers,onecolumn]{revtex4}
\usepackage{graphicx}
\usepackage{amsfonts}
\usepackage{amssymb}
\usepackage{amsmath}
\usepackage{natbib}
\usepackage{dcolumn}
\usepackage{bm} 
\newcommand{\bwt}{\begin{widetext}}
\newcommand{\ewt}{\end{widetext}}
\newcommand{\beq}{\begin{equation}}
\newcommand{\eeq}{\end{equation}}
\newcommand{\bea}{\begin{eqnarray}}
\newcommand{\eea}{\end{eqnarray}}
\begin{document}
\title{Spinodal instabilities and the distillation effect in nuclear matter under strong magnetic fields.}

\author{A. Rabhi}
\email{rabhi@teor.fis.uc.pt}
\affiliation{Centro de F\' {\i}sica Te\'orica, Department of Physics, University of Coimbra, 3004-516 Coimbra, Portugal} 
\affiliation{Laboratoire de Physique de la Mati\`ere Condens\'ee,
Facult\'e des Sciences de Tunis, Campus Universitaire, Le Belv\'ed\`ere-1060, Tunisia}
\author{C.~Provid\^encia}
\email{cp@teor.fis.uc.pt}
\affiliation{Centro de F\' {\i}sica Te\'orica, Department of Physics, University of Coimbra, 3004-516 Coimbra, Portugal} 
\author{J.~Da~Provid\^encia}
\email{providencia@teor.fis.uc.pt}
\affiliation{Centro de F\' {\i}sica Te\'orica, Department of Physics, University of Coimbra, 3004-516 Coimbra, Portugal} 

\date{\today}    
\begin{abstract}
We study the effect of strong magnetic fields, of the order of
$10^{18}-10^{19}$ G,  on the instability region of nuclear matter at subsaturation densities. 
Relativistic nuclear models both with constant couplings and
with density dependent parameters are considered.  It is shown that a strong magnetic
field can have large effects on the instability regions giving rise to
bands of instability and wider unstable regions. As a consequence we predict
larger transition densities at the inner edge of the crust of compact stars
with strong magnetic field. The direction of instability gives rise to a very strong 
distillation effect if the last Landau level is only partially filled. However, 
for almost completed Landau levels an anti-distillation effect may occur. 
 
\end{abstract}
\maketitle

\section{Introduction}

Recent investigations seem to show that soft $\gamma$-ray repeaters and some
anomalous X-ray pulsars are neutron stars which may have surface magnetic fields larger that
$10^{15}$ G~\cite{duncan,usov,pacz}, the so called magnetars.  Until recently, the strongest 
estimated magnetic field is of the order of $B=2\times 10^{15}$ G and was detected 
in a quite young star, SGR 1806-20~\cite{sgr}. According to~\cite{kouve} a fraction as high as 
$10\%$ of the neutron star population could be magnetars. 

The effect of the strong magnetic fields on the equation of state (EOS) of
stellar matter in neutron stars has been
studied both at low densities below neutron drip, of interest for the study of 
the outer crust of neutron stars~\cite{low}, and at  high densities, of interest for the
study of the interior of compact stars~\cite{chakrabarty96,broderick}. In
this last case field-theoretical descriptions based on the non-linear 
Walecka model (NLWM)~\cite{bb}  were used and several
parametrizations compared. It was shown that they have an overall similar
behavior.
It was recently shown in~\cite{aziz08} that the EOS at subsaturation densities, including densities
of the order of the densities at the
inner edge of the crust of a compact star, was particularly affected by fields of
the order of 10$^{18}$ G. 

An important characteristic of the nuclear matter is the appearance of a
liquid-gas phase transition at subsaturation densities. The role of the
isospin is of particular importance. Indeed, since nuclear matter is composed
of two different fluids, namely protons and neutrons, the liquid-gas phase
transition can lead to an isospin distillation phenomenon~\cite{chomaz}. 
 The region of 
instability is determined by the spinodal curve. Due to the symmetry energy, the EOS of $\beta$-equilibrium of magnetic free nuclear matter is thermodynamically stable. The stability of the EOS is determined from the curvature of the free-energy: a positive curvature corresponds to thermodynamically stable matter.

If we
consider stellar matter at very low densities, nuclei in matter are expected
to form a Coulomb lattice embedded in the neutron-electron sea that
minimizes the Coulomb interaction energy. With an increase of the density,
nuclear "pasta" structures emerge~\cite{rpw83}. The existence of pasta phases
may modify some important processes by
changing the hydrodynamic properties and the neutrino opacity in supernova
matter and in the matter of newly born neutron stars~\cite{hor04}. Also, the
pasta phases may influence neutron star quakes and pulsar glitches via the
change of mechanical properties of the crust matter~\cite{quakes}. It is
therefore important to study how the magnetic field could affect the extension
of the pasta phase and the isospin distillation effect.

 In fact, sufficiently strong magnetic field affect the extension of the unstable region.
In order to
have a better understanding of the effect of the magnetic field on the
instabilities of nuclear matter at subsaturation densities we study in the
present work the effect of a strong magnetic field on the thermodynamical
spinodal instabilities obtained from the free energy curvature matrix~\cite{umodes06,abmp06}.
Recently, it was shown that the magnetic field and Joule heating have 
the important effect of maintaining compact stars warm for a
longer time~\cite{pons08}. This kind of simulations need the EOS of the crust. It is,
therefore, important to make a study that shows when should the magnetic field 
be taken explicitly into account in the EOS of the crust. An unstable
region in a wider density range  will correspond to a larger crust and the
properties of the star depending on the crust will be affected. 

In the present paper, we will consider two  relativistic effective approaches: a
NLWM, TM1~\cite{tm1}, with constant coupling parameters,
and a density dependent relativistic hadronic (DDRH) model TW~\cite{tw} 
with density-dependent coupling parameters. DDRH models  seem to give results closer Skyrme interactions  than NLWM,
 at subsaturation densities~\cite{camille08}.

In section II we make a brief review of the models, the EOS under the effect of
a magnetic field and  the stability conditions.  Results are discussed in section III and conclusions are drawn
in the last section. 

\section{The formalism}

\subsection{EOS of nuclear matter under a strong magnetic field}

In the present section we make a short review of
the field-theoretical approach used to obtain the EOS of nuclear matter. Within this approach, the baryons
interact via the exchange of $\sigma$, $\omega$ and $\rho$ mesons in the presence of a
uniform magnetic field $B$ along the $z$-axis.  We start from the Lagrangian
density of TW~\cite{fuchs,tw} model
\beq
{\cal L}= \sum_{b=n, p}{\cal L}_{b} + {\cal L}_{m}.
\label{lan}
\eeq
The baryon ($b$=$n$, $p$) and meson ($\sigma$, $\omega$ and $\rho$) Lagrangians are given by
\bwt
\bea
{\cal L}_{b}&=&\bar{\Psi}_{b}\left(i\gamma_{\mu}\partial^{\mu}-q_{b}\gamma_{\mu}A^{\mu}- 
m_{b}+\Gamma_{\sigma}\sigma
-\Gamma_{\omega}\gamma_{\mu}\omega^{\mu}-\frac{1}{2}\Gamma_{\rho}\tau_{3 b}\gamma_{\mu}\rho^{\mu}
-\frac{1}{2}\mu_{N}\kappa_{b}\sigma_{\mu \nu} F^{\mu \nu}\right )\Psi_{b}, \cr
{\cal L}_{m}&=&\frac{1}{2}\partial_{\mu}\sigma \partial^{\mu}\sigma
-\frac{1}{2}m^{2}_{\sigma}\sigma^{2}
+\frac{1}{2}m^{2}_{\omega}\omega_{\mu}\omega^{\mu}
-\frac{1}{4}\Omega^{\mu \nu} \Omega_{\mu \nu}  \cr
&-&\frac{1}{4} F^{\mu \nu}F_{\mu \nu}
+\frac{1}{2}m^{2}_{\rho}\rho_{\mu}\rho^{\mu}-\frac{1}{4}  P^{\mu \nu}P_{\mu \nu},
\label{lagran}
\eea
\ewt
respectively, where $\Psi_{b}$ are the baryon Dirac fields. The nucleon mass and isospin projection for the protons and neutrons are denoted by $m_{b}$  and $\tau_{3 b}=\pm 1$, respectively. The mesonic and electromagnetic field strength tensors are given by their usual expressions: $\Omega_{\mu \nu}=\partial_{\mu}\omega_{\nu}-\partial_{\nu}\omega_{\mu}$, $P_{\mu 
\nu}=\partial_{\mu}\rho_{\nu}-\partial_{\nu}\rho_{\mu}$, and  $F_{\mu
\nu}=\partial_{\mu}A_{\nu}-\partial_{\nu}A_{\mu}$. The nucleon anomalous
magnetic moments are introduced via the coupling of the baryons to the
electromagnetic field tensor with $\sigma_{\mu
  \nu}=\frac{i}{2}\left[\gamma_{\mu},  \gamma_{\nu}\right] $ and strength
$\kappa_{b}$ with $\kappa_{n}=g_n/2=-1.91315$ for the neutron and
$\kappa_{p}=(g_p/2-1)=1.79285$ for the proton, respectively, and where $g_i$
are the Land\'e $g$ factors for the particle $i$ ($g_p=5.5856912$ and
$g_n=-3.8260837$), and $\mu_N$ is the nuclear magneton. The electromagnetic field is assumed to be externally generated (and thus has no associated field equation), and only frozen-field configurations will be considered.  The electromagnetic couplings are denoted by $q$.
The parameters of the model are the nucleon mass $m_b=939$~MeV, the
masses of mesons $m_\sigma$, $m_\omega$ and $m_\rho$ and the density
dependent coupling parameters $\Gamma$ which are adjusted in order to
reproduce some of the nuclear matter bulk properties and  DBHF (Dirac
Brueckner Hartree-Fock) calculations~\cite{dbhf}, using the following parametrization
\beq
\Gamma_{i}(\rho)=\Gamma_{i}(\rho_{sat})f_{i}(x),\quad i=\sigma, \omega, \rho
\label{gam1}
\eeq
where $x={\rho}/{\rho_{sat}}$, with
\beq
f_{i}(x)=a_{i}\frac{1+b_{i}\left(x+d_{i}\right)^{2}}{1+c_{i}\left(x+d_{i}\right)^{2}},
\quad i=\sigma, \omega,
\label{gam2}
\eeq
and,
\beq
f_{\rho}(x)=\exp\left[ -a_{\rho}(x-1)\right], 
\eeq
with the values of the parameters $m_i$, $\Gamma_{i}$, $a_{i}$, $b_{i}$,
$c_{i}$ and $d_{i}$, $i=\sigma, \omega, \rho$ 
given in Table~\ref{table1} for TW model~\cite{tw}. Other possibilities for these parameters are also found in the
literature~\cite{hbt}. 

\begin{figure}[htb]
\vspace{1.5cm}
\centering
\includegraphics[width=0.7\linewidth,angle=0]{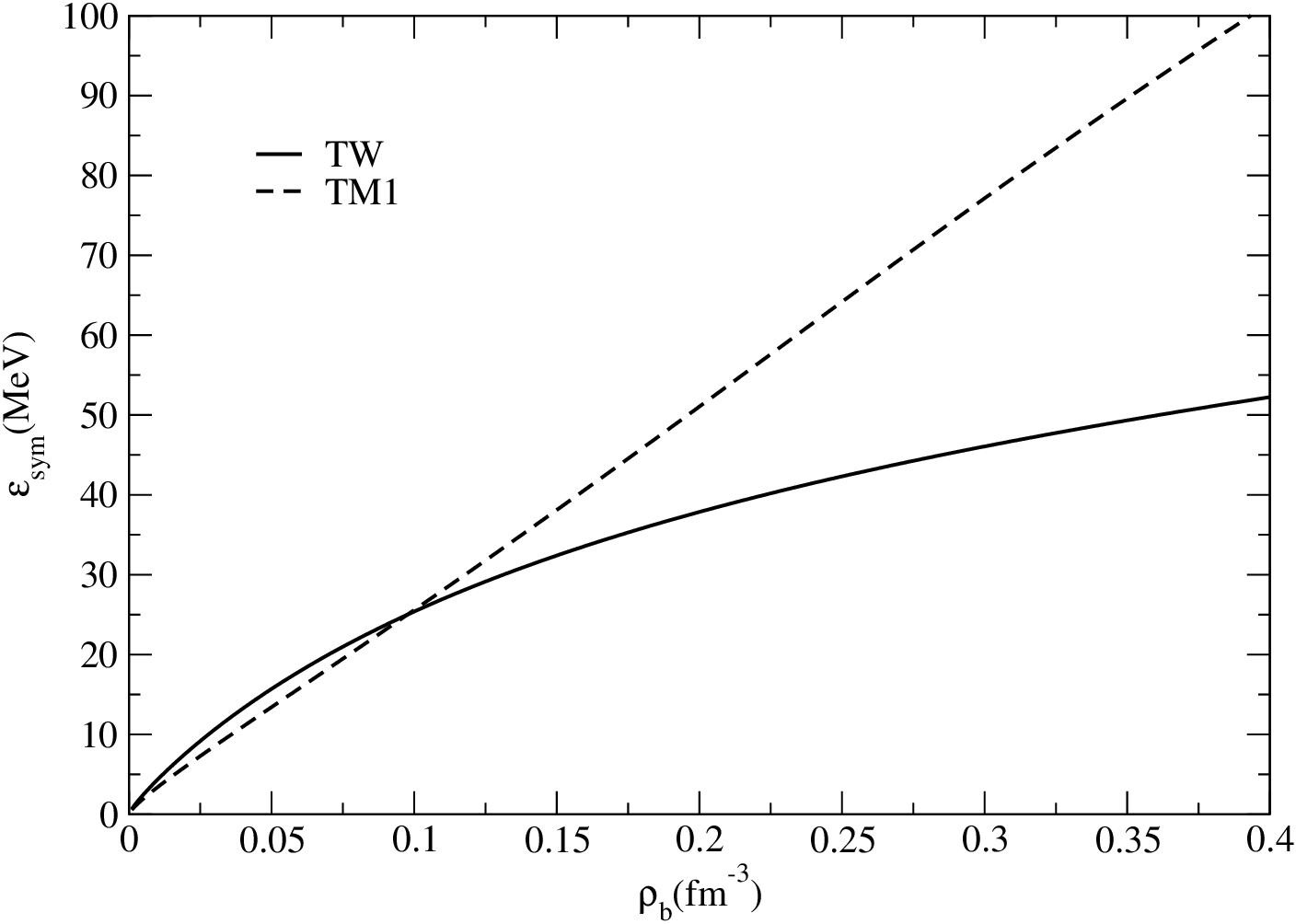}
\caption{Symmetry energy for all models used in the present work. }
\label{esym}
\end{figure}

The symmetry energy and its first and second derivatives are important to understand the instability 
region.
NLWM models become very stiff above saturation densities while DDRH models have 
a softer behavior. On the other hand, at subsaturation densities DDRH models
have larger symmetry energies and  a larger extension of the
instability region for very asymmetric matter. In Fig. \ref{esym} we compare
the symmetry energy of the models we will consider in the present study: TM1
and TW. As expected TM1 has a smaller symmetry energy at subsaturation
densities and a larger one above the saturation density.

\begin{table}[Htb]
\begin{tabular}{ccccccc}
\hline
\hline
i &$m_{i}$(MeV)& $\Gamma_{i}$ & $a_{i}$ & $b_{i}$ & $c_{i}$ & $d_{i}$  \\
\hline
$\sigma$   & 550 &10.72854 &1.365469 &0.226061 &0.409704&0.901995\\
$\omega$ &783 &13.29015 & 1.402488&0.172577 &0.344293&0.983955 \\
$\rho$       &763 & 7.32196 & 0.515 & & & \\
\hline
\hline
\end{tabular}
\caption{Parameters of the TW model.}
\label{table1}
\end{table} 

Notice that in the DDRHM  the nonlinear meson terms are not present, in contrast
with the usual NLWM. For TM1 model we add
to the Langrangian density, Eq. (\ref{lagran}) , with $g_i=\Gamma_i$,
$${\cal L}=\frac{1}{3}bm_n(g_{\sigma}\sigma)^3+\frac{1}{4}c(g_{\sigma}\sigma)^4+\frac{1}{4!}\xi g^4_\omega(\omega_{\mu}\omega^{\mu} )^2. $$
The coupling parameters are constant and given in~\cite{tm1}.

From now we take the standard mean-field theory (MFT) approach and display
only some of the equations. A complete set of equations and description of the
method can be found in the literature (\textit{e.g.}, Ref.~\cite{yuan,
  broderick}). For  the description of the system, we need the baryonic
density, the energy density of nuclear  matter, and the pressure. The energy density of nuclear
matter is given by 
\beq
\varepsilon=\sum_{b=p,n} \varepsilon_{b}+\frac{1}
{2}m^{2}_{\sigma}\sigma^{2}+\frac{1}{2}m^{2}_{\omega}\omega^{2}_{0}+\frac{1}{2}m^{2}_{\rho}\rho^{2}_{0}
\label{ener}
\eeq 
where the energy densities of nucleons have the following forms
\bea
\varepsilon_{p}&=&\frac{q_{p}B}{4\pi^ {2}}\sum_{\nu, s}\left[k^{p}_{F,\nu,s}E^{p}_{F}
+\left(\sqrt{m^{* 2}+2\nu q_{p}B}-s\mu_{N}\kappa_{p}B\right) ^{2} 
\ln\left|\frac{k^{p}_{F,\nu,s}+E^{p}_{F}}{\sqrt{m^{* 2}+2\nu q_{p}B}-s\mu_{N}\kappa_{p}B} \right|\right] , \cr
\varepsilon_{n}&=&\frac{1}{4\pi^ {2}}\sum_{s}\bigg[\frac{1}{2}k^{n}_{F, s}E^{n 3}_{F}-\frac{2}
{3}s\mu_{N}\kappa_{n} B E^{n 3}_{F}\left(\arcsin\left(\frac{\bar{m}_{n}}{E^{n}_{F}} \right)-\frac{\pi}
{2}\right)-\left(\frac{1}{3}s\mu_{N}\kappa_{n} B +\frac{1}{4}\bar{m}_{n}\right) \cr
&&\left(\bar{m}_{n}k^{n}_{F, s}E^{n}_{F}+\bar{m}^{3}_{n}\ln\left|\frac{k^{n}_{F,s}+E^{n}_{F}}{\bar{m}_{n}} 
\right|\right) \bigg].
\eea
For the neutrons we have introduced
\beq
\bar{m}_{n}=m^{*}-s\mu_{N}\kappa_{n}B,
\label{barm}
\eeq
where the effective baryon masses are given by 
\beq
m^{*}=m-\Gamma_{\sigma}\sigma.
\eeq 

The pressure of the system is obtained from the expression
\beq
P_{m}=\sum_{b=n,p}\mu_{b}\rho_{b}-\varepsilon.
\label{press}
\eeq
  
The energy spectra for protons are neutrons are given by
\bea
E^{p}_{\nu, s}&=& \sqrt{k^{2}_{z}+\left(\sqrt{m^{* 2}+2\nu q_{p}B}-s\mu_{N}\kappa_{p}B \right) 
^{2}}+\Gamma_{\omega} \omega^{0}+\frac{1}{2}\Gamma_{\rho}\rho^{0}+\Sigma^{R}_{0}, \label{enspc1}\\
E^{n}_{s}&=& \sqrt{k^{2}_{z}+\left(\sqrt{m^{* 2}+k^{2}_{x}+k^{2}_{y}}-s\mu_{N}\kappa_{n}B 
\right)^{2}}+\Gamma_{\omega} \omega^{0}-\frac{1}{2}\Gamma_{\rho}\rho^{0}+\Sigma^{R}_{0},
\label{enspc2} 
\eea
respectively, where $\nu=n+\frac{1}{2}-sign(q)\frac{s}{2}=0, 1, 2, \ldots$ enumerates the Landau levels of the fermions with electric charge $q$, the quantum number $s$ is $+1$ for spin up and $-1$ for spin down
cases, and the rearrangement term is given by 
\beq
\Sigma^{R}_{0}=\frac{\partial \Gamma_{\omega}}{\partial \rho}\rho_b\omega_{0}+\frac{\partial \Gamma_{\rho}}{\partial 
\rho}\rho_{3}\frac{\rho_{0}}{2}-\frac{\partial \Gamma_{\sigma}}{\partial \rho}\rho^s\sigma,
\eeq
where $\rho^s=\rho^s_p+\rho^s_n$ and $\rho_b=\rho_p+\rho_n$, with the expressions of the scalar and vector densities for protons and neutrons given by~\cite{broderick}
\bea
\rho^{s}_{p}&=&\frac{q_{p}Bm^{*}}{2\pi^{2}}\sum_{\nu=0}^{\nu_{\mbox{\small max}}}\sum_{s}\frac{\sqrt{m^{* 2}+2\nu 
q_{p}B}-s\mu_{N}\kappa_{p}B}{\sqrt{m^{* 2}+2\nu q_{p}B}}\ln\left|\frac{k^{p}_{F,\nu,s}+E^{p}_{F}}
{\sqrt{m^{* 2}+2\nu q_{p}B}-s\mu_{N}\kappa_{p}B} \right|, \cr
\rho^{s}_{n}&=&\frac{m^{*}}{4\pi^{2}}\sum_{s} \left[E^ {n}_{F}k^{n}_{F, s}-\bar{m}^{2}_{n}\ln\left|
\frac{k^{n}_{F,s}+E^{n}_{F}}{\bar{m}_{n}} \right|\right],  \cr
\rho_{p}&=&\frac{q_{p}B}{2\pi^{2}}\sum_{\nu, s}k^{p}_{F,\nu,s},  \cr
\rho_{n}&=&\frac{1}{2\pi^{2}}\sum_{s}\left[ \frac{1}{3}\left(k^{n}_{F, s}\right) ^{3}-\frac{1}
{2}s\mu_{N}\kappa_{n}B\left(\bar{m}_{n}k^{n}_{F,s}+E^{n 2}_{F}\left(\arcsin\left( \frac{\bar{m}_{n}}
{E^{n}_{F}}\right) -\frac{\pi}{2} \right)  \right) \right] 
\eea
where $k^{p}_{F, \nu, s}$ and $ k^{n}_{F, s}$ are the Fermi momenta of protons and neutrons which are related to the Fermi energies $E^{p}_{F}$ and $E^{n}_{F}$ through
\bea
k^{p 2}_{F,\nu,s}&=&E^{p 2}_{F}-\left[\sqrt{m^{* 2}+2\nu q_{p}B}-s\mu_{N}\kappa_{p}B\right] ^{2} \cr
k^{n 2}_{F,s}&=&E^{n 2}_{F}-\bar{m}^{2}_{n}.
\eea

The chemical potentials of nucleons within TW are given  by
\bea
\mu_{p}&=& E^{p}_{F}+\Gamma_{\omega}\omega^{0}+\frac{1}{2}\Gamma_{\rho}\rho^{0}
+\Sigma^{R}_{0} \cr
\mu_{n}&=& E^{n}_{F}+\Gamma_{\omega}\omega^{0}-\frac{1}{2}\Gamma_{\rho}\rho^{0}
+\Sigma^{R}_{0}.
\label{mu}
\eea
For TM1 we have similar expressions with the last term, the rearrangement
term, equal to zero. 

\subsection{Stability conditions for nuclear matter}

At subsaturation densities nuclear matter has a liquid-gas phase transition
and homogeneous matter is not stable within a given range of densities. 
The stability conditions for asymmetric nuclear matter, keeping volume and temperature constant, 
are obtained from the free energy density $\cal F$, imposing that this
function is a convex function of the densities $\rho_p$ and $\rho_n$. For
stable homogeneous matter, the symmetric matrix with the elements~\cite{Bar03,  marg03}, 
\beq
{\cal F}_{ij}=\left( \frac{\partial^{2} {\cal F}}{\partial \rho_{i}\partial\rho_{j}}\right) _{T},
\eeq
known as stability matrix,  must be positive. This corresponds to imposing that the trace and the
determinant of ${\cal F}_{ij}$ are positive. In terms of the proton and
neutron chemical potentials $\mu_i$, the stability matrix is given by
\beq
{\cal F}=
\begin{pmatrix}
\displaystyle\frac{\partial\mu_{n}}{\partial \rho_{n}} 
&\displaystyle\frac{\partial\mu_{n}}{\partial \rho_{p}} \\
\displaystyle\frac{\partial\mu_{p}}{\partial \rho_{n}}
&\displaystyle\frac{\partial\mu_{p}}{\partial \rho_{p}}
\label{cur}
\end{pmatrix},
\eeq
with $\mu_{i}=\frac{\partial{\cal F}}{\partial \rho_{i}}|_{T,\rho_{j\neq i}} $.

For nuclear matter,
the largest eigenvalue of the stability matrix is always positive and the
other becomes negative at subsaturation densities.
We define the thermodynamical spinodal at T=0 as the curve on the  $\rho_n$,
$\rho_p$ plane defined by the points  for which the determinant of $ {\cal  F}_{ij}$ is zero; 
that is, the smallest  eigenvalue is zero. Inside the region limited by the thermodynamical
spinodal  the smallest eigenvalue of  ${\cal F}_{ij}$ is negative and nuclear matter is 
unstable~\cite{marg03}. At $T=0$, $\cal F$ is equal to the energy density defined 
by Eq. (\ref{ener}). The eigenvalues of the stability matrix are given by
\beq
\lambda_{\pm}=\frac{1}{2}\left(\hbox{Tr}({\cal F})\pm\sqrt{\hbox{Tr}({\cal F})^2-4 \hbox{Det}({\cal F})}\right).
\label{lambd}
\eeq
The stability condition requires that they are both positive. When one curvature becomes negative the system is thermodynamically unstable and can decrease its free energy by going in the instability direction \cite{marg03}, defined by the direction of the eigenvector associated to the negative eigenvalue.
The eigenvectors $\delta {\bf \rho}_{\pm}$ of the stability matrix are given by 
\beq
\frac{\delta \rho^{\pm}_{i}}{\delta \rho^{\pm}_{j}}=\frac{\lambda_{\pm}-{\cal F}_{jj}}{{\cal F}_{ji}}, \: i, j = p, n.
\eeq
In the following we study the direction of instability inside the spinodal
section for both  models considered.

\subsection{Stability conditions  for $npe$ matter}
Stellar matter at low densities is formed by protons, neutrons and
electrons in equilibrium with respect to weak interaction processes. Until now we have considered the subsaturation instability region
of neutron-proton ($np$) nuclear matter.  In this section we investigate the
effect of the inclusion of electrons on the stability conditions of nuclear matter when electrons are
included. In particular, we will calculate the thermodynamical spinodal
sections for $npe$ (neutron-proton-electron) neutral matter. Since matter is neutral the proton and electron densities must be equal, i.e. $\rho_p=\rho_e$.

Electrons are included in a minimal way in the system, and are   described by the following the Lagrangian density 
\beq
{\cal L}_{e} = \bar{\psi}_{e}\left(i\gamma_{\mu}\partial^{\mu}-q_{e}\gamma_{\mu}A^{\mu}
-m_{e}\right )\psi_{e},
\label{lagran}
\eeq
where $\psi_{l} $ are the electron Dirac fields and $m_e=0.511 \hbox{ MeV}$. 

Including the electrons, the stability matrix  (\ref{cur}) becomes
\beq
{\cal F}=
\begin{pmatrix}
\displaystyle\frac{\partial\mu_{n}}{\partial \rho_{n}} &\displaystyle\frac{\partial\mu_{n}}{\partial \rho_{p}} \\
\displaystyle\frac{\partial\mu_{p}}{\partial \rho_{n}}
&\displaystyle\frac{\partial\left(\mu_{p}+\mu_{e}\right)}{\partial \rho_{p}}
\end{pmatrix},
\label{cur1}
\eeq
and the stability conditions are equivalent to the ones indicated in the previous subsection: the trace and the determinant of ${\cal F}$ must be positive.

The electron density is given by
\beq
\rho_e=\frac{|q_{e}|B}{2\pi^{2}}\sum_{\nu, s}k^{e}_{F, \nu, s}
\eeq
where $k^{e}_{F, \nu, s}$ is  the electron Fermi momentum related to the Fermi energy $E^{e}_{F}$ by
\beq
k^{e 2}_{F,\nu,s}=E^{e 2}_{F}-\left(m^{2}_{e}+2\nu |q_{e}| B\right).
\eeq
For $npe$ neutral matter
$${\cal F}=\epsilon+ \epsilon_e$$
where $\epsilon$ was defined in (\ref{ener}) and
the electron contribution is given by
\beq
\varepsilon_{e}=\frac{|q_{e}|B}{4\pi^ {2}}\sum_{\nu, s}\left[k^{l}_{F,\nu,s}E^{e}_{F}
+\left(m^{2}_{e}+2\nu |q_{e}|B\right) 
\ln\left|\frac{k^{e}_{F,\nu,s}+E^{e}_{F}}{\sqrt{m^{2}_{e}+2\nu |q_{e}| B}} \right|\right] .
\eeq
We next discuss $np$  nuclear matter and  $npe$ neutral matter. $\beta$-equilibrium stellar matter is a particular case of  $npe$ neutral matter, with the proton and electron fractions defined by chemical equilibrium conditions, namely
$$\mu_p=\mu_n-\mu_e,$$
with $\mu_p$ and $\mu_n$ defined in (\ref{mu}) and $\mu_e=E^{e }_{F}$.
 
\section{Results and discussion}

In the present section we first show the dependence of the spinodal section on  the magnetic field, 
both for neutron-proton  ($np$) and neutron-proton-electron ($npe$) matter. 
From the crossing of the  $\beta$-equilibrium  EOS of stellar matter with the thermodynamical spinodal we make a prevision of the transition density and transition pressure at the inner edge of the crust of a compact star. 

For a zero magnetic field,  the direction of instability of nuclear matter gives rise to a 
distillation effect, corresponding to the formation of droplets of  dense matter with low isospin asymmetry
in a background of a neutron gas with a small fraction of protons. This effect has been observed experimentally in heavy-ion reactions \cite{chomaz}. Therefore, 
we also discuss the effect of the magnetic field on the direction of instability, namely in which way it affects the distillation effect.

\subsection{Spinodal section $np$ matter}

\begin{figure}[htb]
\centering
\includegraphics[width=0.7\linewidth,angle=0]{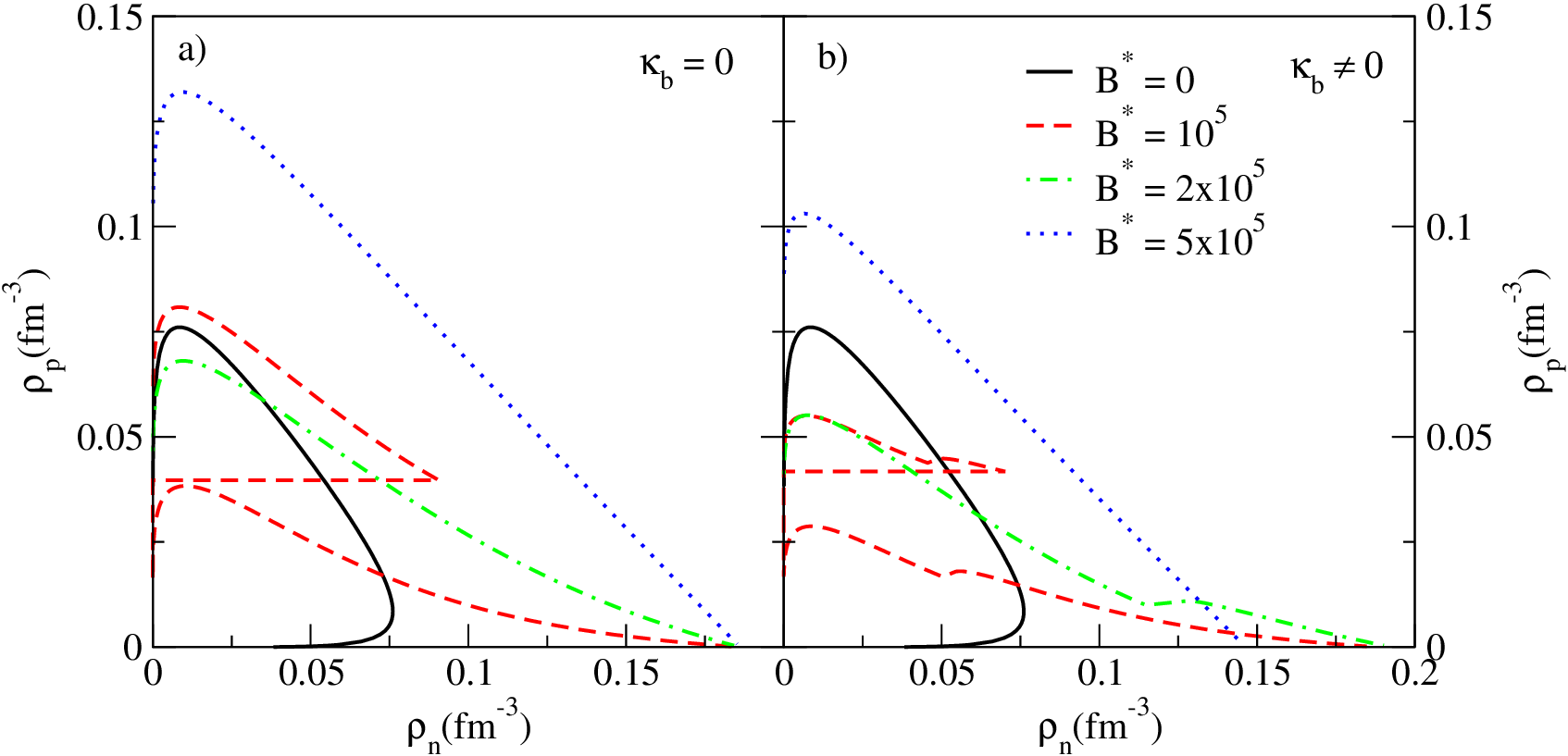}
\caption{(Color online) Spinodal section on the  $\rho_p$, $\rho_n$
  plane for TM1 at $T = 0 \hbox{ MeV}$ and for severals values of
  magnetic fields, $B=B^*\, B_e$, a)  without  and b) with  AMM. For $B^*=10^5$ the spinodal section is formed by two separate regions.}
\label{spztm1npn}
\end{figure}

\begin{figure}[htb]
\vspace{1.5cm}
\centering
\includegraphics[width=0.7\linewidth,angle=0]{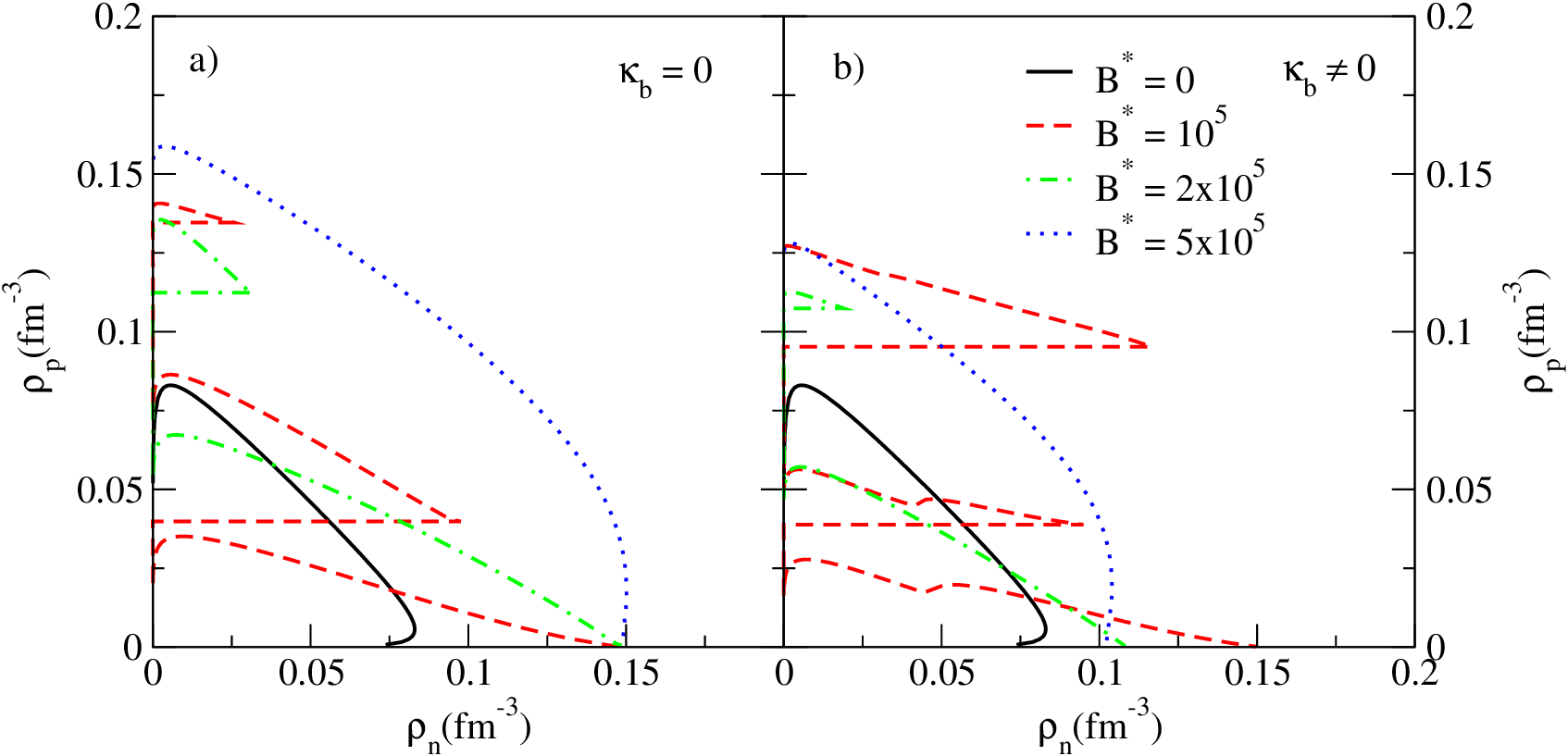}
\caption{(Color online) Spinodal section in terms of $\rho_p$ versus $\rho_n$
for TW at $T = 0 \hbox{ MeV}$ and for severals values of magnetic fields
a) without  and b) with  AMM. For $B^*=10^5$ and $2\times 10^5$ the spinodal section has respectively three and two separated parts.}
\label{spztwnp}
\end{figure}

We will first consider $np$ nuclear  matter and determine the instability region limited by the spinodal surface.
In Fig.~\ref{spztm1npn} and Fig.~\ref{spztwnp} we show the spinodal sections obtained making $\lambda_-=0$, where $\lambda_-$ was defined in Eq.  (\ref{lambd}),  on
the ($\rho_p$, $\rho_n$) plane for TM1  and TW and  for several values of the
magnetic field.
We define the magnetic field in units of the critical  field $B^c_e=4.414 \times 10^{13}$~G, so that  $B=B^* \, B^c_e$. For a field with the intensity $B^c_e$ the electron cyclotron energy is equal to the electron mass. 

 We present the numerical results both not including and including the contribution of the anomalous magnetic moment (AMM).  In all figures where both cases are considered we 
show on the left panel the results without the magnetic field   and on the right panel the results
including  the AMM. 

The magnetic
field has a strong effect not only on the size, giving rise to larger
instability regions, but also on the shape of the
instability zones which is not symmetric with respect to the $\rho_p=\rho_n$ line, contrary to $np$ matter for $B=0$. 
The increase of the instability region is due to 
Landau quantization which softens the EOS. 
 For magnetic fields $B^{*}>2\times10^{5}$, the
protons are totally polarized for all the values of the densities
considered and the size of the spinodal zone is larger than
the one obtained for $B=0$.
Including AMM decreases the spinodal region with respect to the results
without AMM for all the values of the magnetic field considered.
This behavior is explained by the extra stiffness that  
the inclusion of AMM brings  into  the system due to neutron spin polarization.
\bigskip

\begin{figure}[htb]
\vspace{1.5cm}
\centering
\includegraphics[width=0.7\linewidth,angle=0]{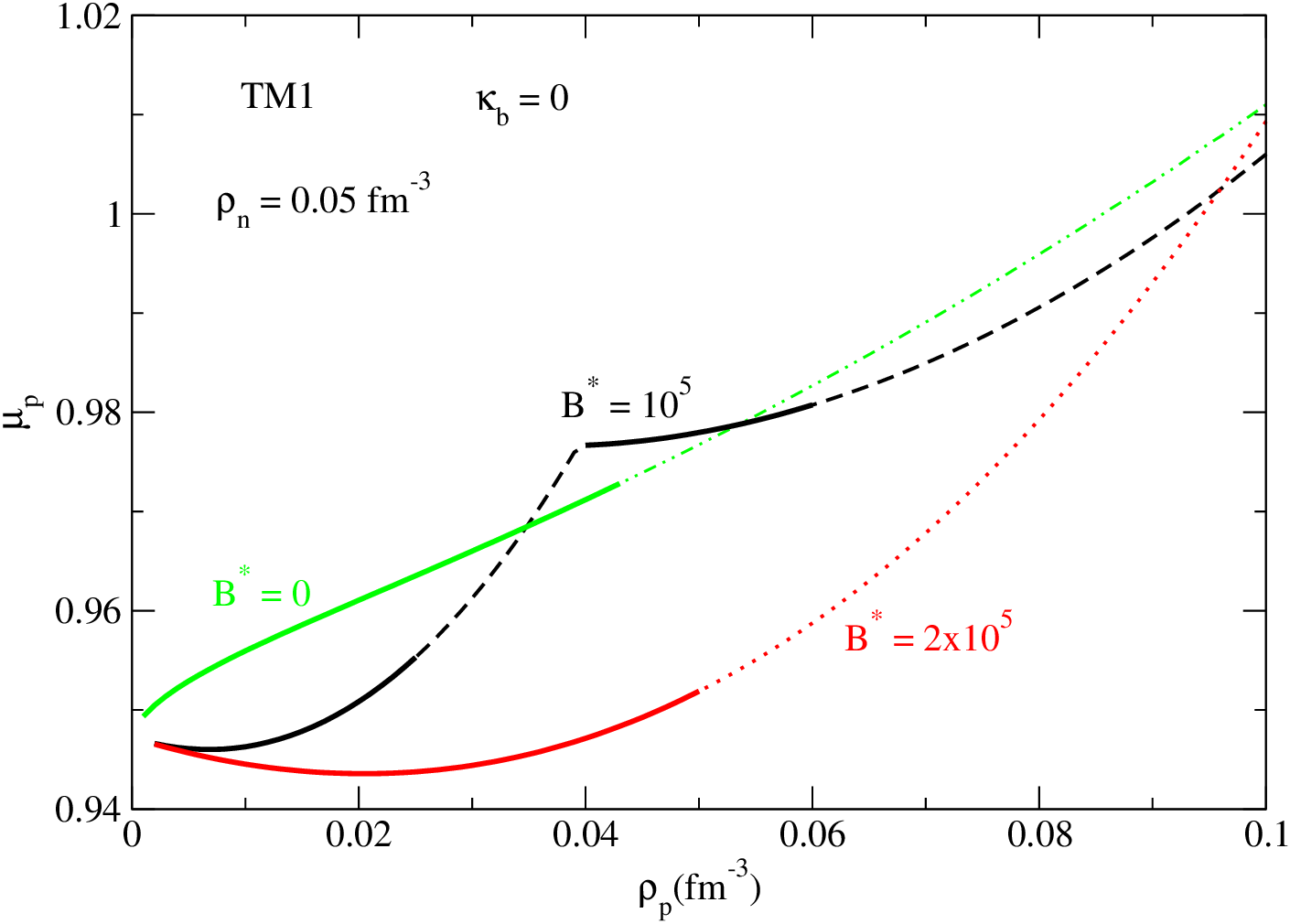}
\caption{(Color online) The proton chemical potential for $B=0$ and for two magnetic field  intensities, $B^*=10^5$ and  $B^*=2\times 10^5$, obtained within the TM1
model for the neutron density $\rho_n=0.05$ fm$^{-3}$ and excluding the
AMM. The thick lines represent the regions of instability.}
\label{chem2}
\end{figure}

The effect of the Landau quantization on the spinodal section  is explicitly seen
 in the spinodal for $B^{*}=10^{5}$ for TM1 [dashed line
Fig. \ref{spztm1npn} a)]. The
spinodal region consists of two separate zones each one corresponding to one
Landau level, the one corresponding to the first Landau level extends to
larger neutron densities than the one corresponding to the second level.
In order to understand this effect,  we plot in Fig. \ref{chem2} the proton chemical potential for $B^*=0$, 
$B^{*}=10^{5}$ and $B^{*}=3\times 10^{5}$. We have  identified the
instable regions with thick lines. It is seen that for the larger
field the proton chemical potential changes smoothly  because for all
the densities shown only the first Landau level (LL) is occupied. At
low densities the chemical potential decreases with density and only
above 0.025 fm$^{-3}$ it starts increasing with density. For
$B^{*}=10^{5}$ the proton chemical potential shows a cusp
corresponding to the end of the first LL and beginning of the second LL. The unstable regions, in this case, correspond to the beginning of each LL when the slope of the chemical potential is smaller. The smaller the magnetic field the larger the number of LL occupied at subsaturation densities and the larger of independent sections which make up the whole spinodal section. For reference we include the chemical potential at $B=0$: it increases smoothly with density with a quite constant slope.

If the AMM is included the instabilities regions are smaller, as discussed above.
The structure (bump) appearing at $\rho_n\sim 0.05$ fm$^{-3}$ is due to the neutron
polarization: below that value of the density the neutrons are totally polarized. 


For the TW model with $B^{*}=2 \times 10^{5}$  and  $10^{5}$, Fig. \ref{spztwnp},  we also get a 
spinodal region formed by several bands, respectively two and three bands. 
An interesting feature of this model, is that the band with the largest Landau level may
extend to larger neutron densities than lower levels. This does not occur for NLWM and 
has to do with the behavior of the symmetry energy which increases in a smoother 
way for DDRHM than for NLWM above $\rho=0.1$ fm$^{-3}$. As a result the slope of the chemical potentials is not so large.

Another feature of the spinodal regions with magnetic field and without AMM, that is
present in both models we have studied, is the extension of the spinodal for
zero proton fraction: in the absence of the magnetic field there is no
instability but the inclusion of the magnetic field changes this behavior: the instability region 
at $\rho_p=0$ extends until a finite  $\rho_n$ value, model dependent, but independent of
$B$. This value is $\sim 0.15$ fm$^{-3}$ for TW and  $\sim 0.186$ fm$^{-3}$ for TM1. 

To understand this behavior seen at zero proton fraction, we consider the TM1 model.  For $\rho_p=0$, we obtain the
corresponding finite value of $\displaystyle\rho_n=\frac{k^{n 3}_ {F}}{3\pi^2}$ from the Fermi neutron momenta $k^{n}_ {F}$ solution of the equation $\hbox{Det}({\cal F})=0$ with $\rho_p=0$. Explicitly, the latter equation can be written as follows 

\bea
\left({\cal A}_{+}-\frac{\cal C}{m^{* 2}} -{\cal D} \right) \left(\frac{\pi^{2}}{E^{n}_{F} k^{n}_{F}}+{\cal A}_{+}-\frac{\cal C}{E^{n 2}_{F}}-{\cal D} \right) -\left({\cal A}_{-}-\frac{\cal C}{m^{*} E^{n }_{F}}-{\cal D} \right)^{2} = 0
\eea
where $\displaystyle{\cal A}_{\pm}=\left(\frac{g_{\omega}}{m_{\omega}}\right)^{2}
\pm\frac{1}{4}\left(\frac{g_{\rho}}{m_{\rho}}\right)^{2}$, $E^{n}_{F}=\sqrt{k^{n 2}_ {F}+m^{* 2}}$ , and $\displaystyle {\cal C}=\left(\frac{g_\sigma}{m_{\sigma}}\right)^2 \frac{m^{* 2}}{\cal K}$ with,
\bea
{\cal K}=1+\left(\frac{g_{\sigma}}{m_{\sigma}}\right)^{2}\left[2 b (g_{\sigma}\sigma)+3c(g_{\sigma}\sigma)^2+\frac{1}{2 \pi^ 2 E^{n}_{F}}\left(k^{n 3}_ {F}+3 m^{* 2} k^{n}_ {F}-3 m^{* 2} E^{n}_{F}\log \left|\frac{k^{n}_ {F}+E^{n}_{F}}{m^{*}}\right|\right) \right] 
\eea
and $\displaystyle{\cal D}=\frac{\frac{1}{2}\xi\left( \frac{g_{\omega}}{m_{\omega}}\right)^{4}\left(g_{\omega}\omega^0 \right)^2}{1+\frac{1}{2}\xi\left( \frac{g_{\omega}}{m_{\omega}}\right)^{2}\left(g_{\omega}\omega^0 \right)^2}.$ \\
This equation is independent of the magnetic field and, therefore, all the
spinodal regions for different magnetic fields, without AMM, in
Figs.~\ref{spztm1npn} and~\ref{spztwnp} have
the same value of $\rho_n$ for $\rho_p=0$. The inclusion of AMM changes this
feature: $\rho_n$ is still finite for $\rho_p=0$ but its value is $B$ dependent.

For models with density dependent couplings the spinodal region 
extends to larger  densities for the larger proton densities
when compared with NLWM.
This is due to the behavior of the symmetry energy: while for
NLWM the symmetry energy increases quite steeply for densities above
saturation densities, DDRH models have a much smoother behavior and the
symmetry energy of these models take much smaller values than NLWM for
densities above $\rho=0.15$ fm$^{-3}$.

\begin{figure}[htb]
\vspace{1.5cm}
\centering
\includegraphics[width=0.7\linewidth,angle=0]{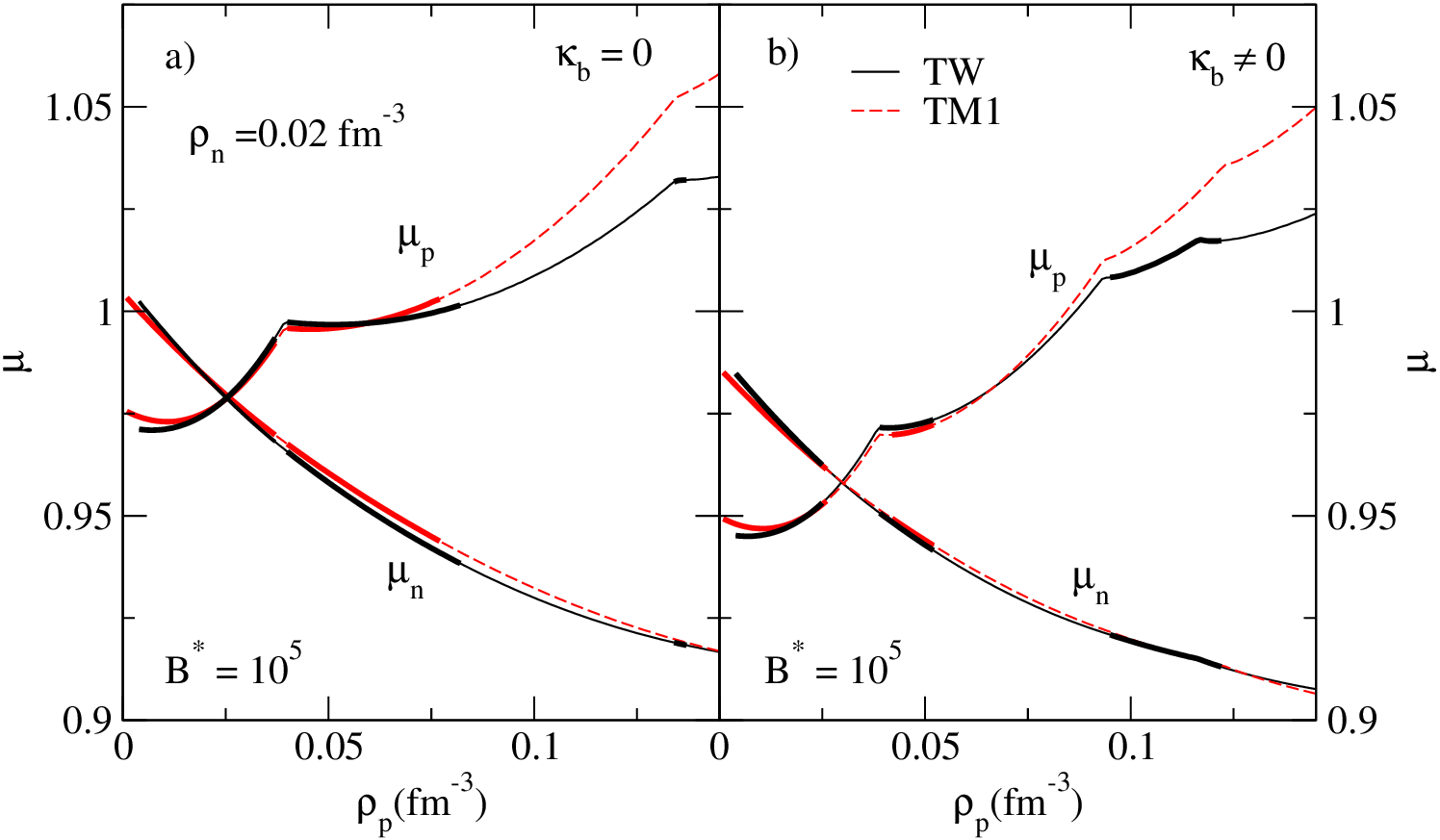}
\caption{(Color online) Proton and neutron chemical potentials  as function of
the proton density for $\rho_n = 0.02$  $\hbox{ fm}^{-3}$ and for $B^*=10^5$
a) without and b) with  AMM for TM1 (dashed line)  and TW (full line).
The thick segments of each curve represent the regions of instability.}
\label{chem}
\end{figure}

\begin{figure}[htb]
\vspace{1.5cm}
\centering
\includegraphics[width=0.7\linewidth,angle=0]{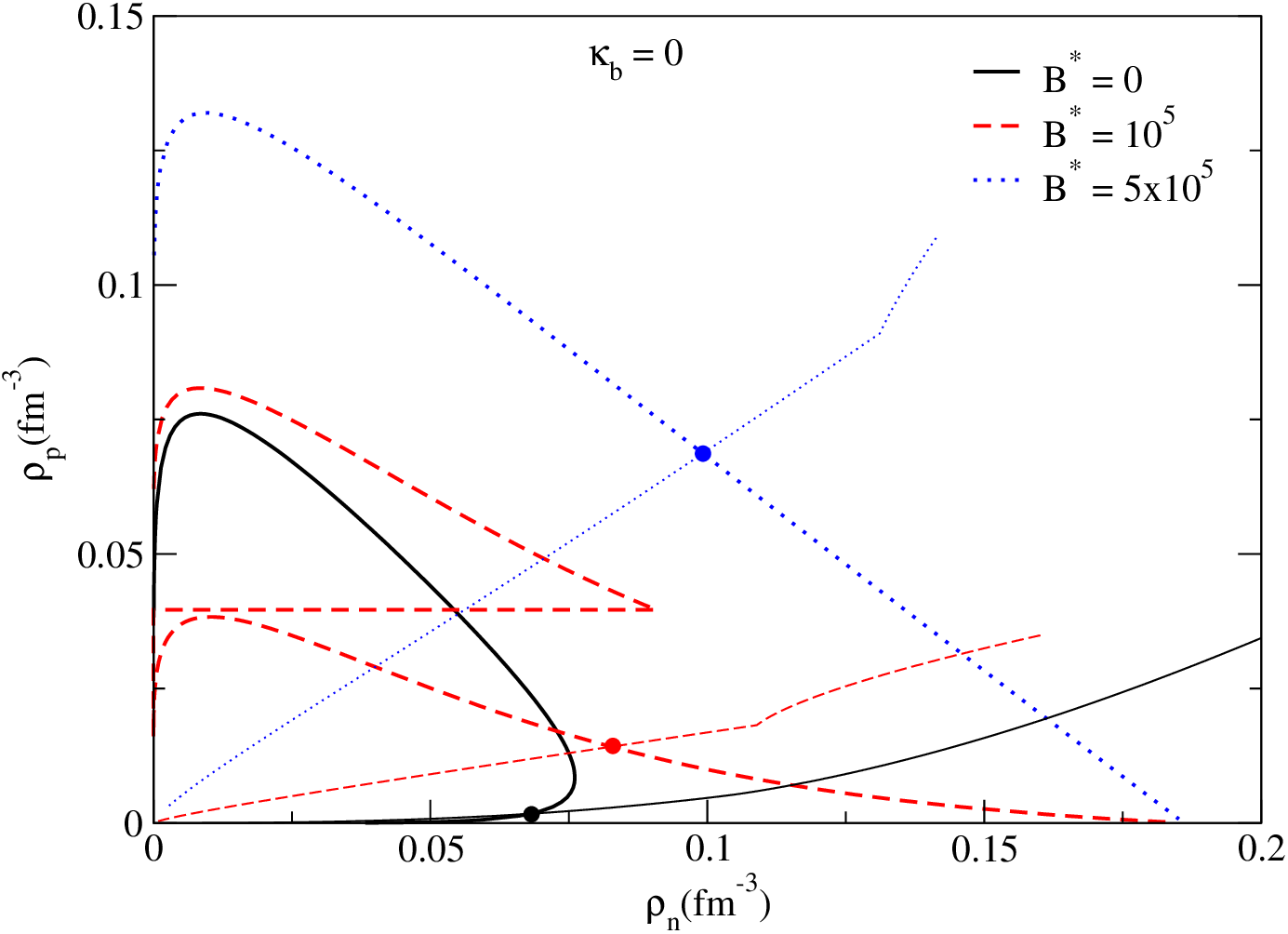}
\caption{(Color online) Spinodal section in the  $\rho_p$,  $\rho_n$ plane for
TM1 at $T = 0 \hbox{ MeV}$ and for severals values of the magnetic field without
AMM. For each value of the magnetic field, it also plotted the EOS of
stellar matter in $\beta$-equilibrium  (thin lines). The crossing of the EOS
with the respective spinodal, large dot in each spinodal, represent the
transition density to continuous matter. }
\label{spztm1npd}
\end{figure}

 In Fig. \ref{chem} we plot $\mu_p$ and  $\mu_n$ for $B^*=10^5$ with
the models TM1 (dashed lines) and TW (full lines). The thick segments of the curves lines identify
the instability regions defined by the two (TM1) and three (TW) bands which form the spinodal.  These curves are obtained for a fixed
neutron density, $\rho_n=0.02$ fm$^{-3}$. It is seen that above  $\rho_p=0.1$
fm$^{-3}$ the proton chemical potential within TW is much softer and this
seems to be the reason for the appearance of the third band in this model.

For each model TM1 and TW we  identified the crossing density,
and corresponding pressure, of the EOS of $\beta$-equilibrium stellar
matter and the corresponding spinodal for each value of the magnetic
field considered.  The EOS of state was obtained considering neutrons,
protons and electrons in $\beta$-equilibrium. In order to illustrate
what was done we represent in Fig.~\ref{spztm1npd} the spinodal
sections obtained within TM1 for $B^*=0, 10^5$ and $5\times10^5$
respectively by full, dashed and dotted thick lines. We include in the
same figure, using thin lines with the same type of curve for each $B$ value, the corresponding EOS of  $\beta$-equilibrium stellar matter. The crossing spinodal-EOS is identified by a big dot. Both the spinodal and the EOS are plotted in the $\rho_p, \rho_n$ plane. 

The crossing density of the EOS with the thermodynamical spinodal gives a 
prevision of the transition density \cite{bao, pasta1} to an homogeneous phase, 
and is always larger than the one obtained from the crossing of the EOS with the
dynamical spinodal for $npe$ matter, which includes the Coulomb
interaction. In \cite{link99} the authors have shown how the transition density
and respective pressure were related to the fraction of the star's
moment of inertia contained in the solid crust, and obtained
 a relation between the radius and mass of compact stars. 

In Tables~\ref{table4} and \ref{table5} the values of the crossing density and
respective pressure are given for stellar matter under different magnetic
fields, respectively without and with the AMM.
The values of the crossing density for $B=0$, 0.069 fm$^{-3}$ for TM1 and  0.085 fm$^{-3}$  for TW,  can be compared with the
corresponding ones obtained from the crossing of the dynamical spinodal with
the EOS Ref. \cite{camille08,pasta1}, respectively 0.06 fm$^{-3}$ for TM1  and 0.075 fm$^{-3}$ for TW. As expected they are a
bit larger, with TW model having a larger crossing density than the other.
The effect of the magnetic field is to increase the values of the crossing
density: at $B^*=10^5$ both models have similar transition densities of the
order of $\sim 0.1$ fm$^{-3}$ corresponding to a much larger pressure for 
TM1 than TW. For  $B^*=3\times 10^5$ the transition densities increase to
$\rho\sim 0.14-0.15$ fm$^{-3}$.

\begin{table*}[t]
\caption{
Predicted density, proton fraction and pressure at the inner edge of the
crust of a compact star at zero temperature, as defined by the crossing
between the thermodynamical instability region of $np$ matter and the
$\beta$-equilibrium EOS for homogeneous, neutrino-free stellar
matter in the $\rho_p,\rho_n$ plane. The AMM is not included.} 
\label{table4}
\begin{ruledtabular}
\begin{tabular}{ccccc}
$B^{*}$  & Models & $\rho^{\hbox{cross}}_b (\hbox{ fm}^{-3}) $ & $Y_p$ & $P_{m}(\hbox{ MeV}\hbox{ fm}^{-3}) $ \\
\hline
  $0$            & TM1 & 0.069509 & 0.024713 & 0.50288 \\
                                & TW  & 0.084955 & 0.036690  & 0.52246 \\
$10^{5}$         & TM1  & 0.097030 & 0.14645 & 0.95944  \\
                                       &  TW  & 0.10099   & 0.14641   & 0.67321  \\
$2\times10^{5}$    & TM1 & 0.12266 & 0.24283 & 1.4008  \\
                                             & TW  & 0.12786  & 0.23599 & 1.0156  \\
$3\times10^{5}$    & TM1 & 0.14085 & 0.31304 & 1.5944 \\
                                             & TW  & 0.15159 & 0.30219 & 1.3795 \\
$5\times10^{5}$    & TM1 & 0.16783 & 0.40921  & 1.6324 \\
                                             & TW &0.19784 & 0.40194 & 2.3310 \\
\end{tabular}
\end{ruledtabular}
\end{table*}

In Table~\ref{table5} we show the same data given in Table~\ref{table4} but
including the AMM in the calculation. The conclusions are similar: the
transition density increases with the increase of the magnitude of the magnetic field but not
so fast. However, the corresponding pressures are larger than before. We conclude that the existence of a strong magnetic field at the crust gives rise to a larger crust.

\begin{table*}[t]
\caption{
Predicted density, proton fraction and pressure at the inner edge of the crust of a compact star at zero temperature, as defined by the crossing between the thermodynamical instability region of $np$ matter and the $\beta$-equilibrium condition for homogeneous, neutrino-free stellar matter. The case where the AMM is  included.}  
\begin{ruledtabular}
\label{table5}
\begin{tabular}{c c c c c}
$B^{*}$ & Models & $\rho^{\hbox{cross}}_b (\hbox{ fm}^{-3}) $ & $Y_p$ & $P_{m}(\hbox{ MeV}\hbox{ fm}^{-3}) $ \\
\hline
           $10^{5}$             & TM1  & 0.086942 & 0.17670  & 1.3801  \\
                                           &TW  & 0.091391 & 0.17829 & 1.2809 \\
 $2\times10^{5}$ &TM1 & 0.096337 & 0.29468 & 1.5373 \\
                                           &TW & 0.092438 & 0.30512 & 1.3549  \\
$3\times10^{5}$ &TM1 & 0.11046 & 0.37135 & 1.7091 \\
                                           &TW  & 0.11251 & 0.38035 & 1.8047 \\
 $5\times10^{5}$ & TM1  & 0.12822 & 0.47093 & 1.6616 \\
                                            & TW    & 0.14880 & 0.48803 & 2.7717 \\
\end{tabular}
\end{ruledtabular}
\end{table*}

\subsection{Spinodal section $npe$ matter}

We have studied the effect of the magnetic field on the instability region of
$np$ matter in the previous section.
For $npe$ matter without magnetic field, NLWM models still present a small
thermodynamical instability region but for DDRHM models there is no instability region \cite{abmp06}. The
incompressibility of the free electron gas is so high that the spinodal
disappears or almost disappears.     

\begin{figure}[htb]
\vspace{1.5cm}
\centering
\includegraphics[width=0.7\linewidth,angle=0]{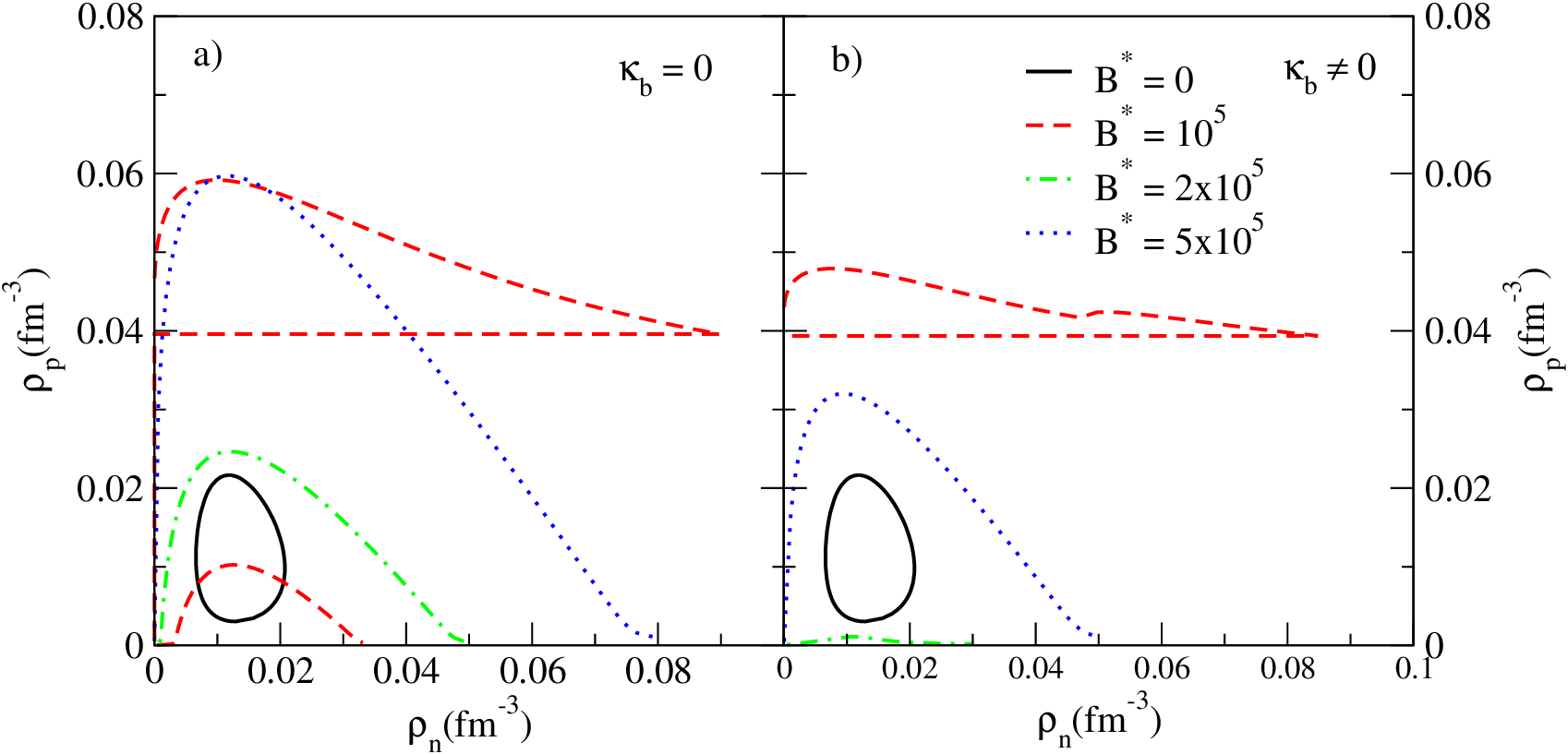}
\caption{(Color online) Spinodal section in terms of $\rho_p$ versus $\rho_n$ 
for TM1 for $npe$ neutral matter at $T = 0 \hbox{ MeV}$ and for severals values of magnetic 
fields (a) without  and (b) with  AMM.}
\label{spin_npe_nlw}
\end{figure}

 In Fig. \ref{spin_npe_nlw} the spinodals for $npe$ matter are shown
 for TM1 and different magnetic fields. 
In fact,  although including electrons, the instability region can become almost as
large as the $B=0$ $np$-spinodal.
This is due both to the Landau quantization of the orbital motion of protons
and electrons: the incompressibility of the electron gas is smaller than the
one of a magnetic free electron gas.

\begin{figure}[htb]
\vspace{1.5cm}
\centering
\includegraphics[width=0.7\linewidth,angle=0]{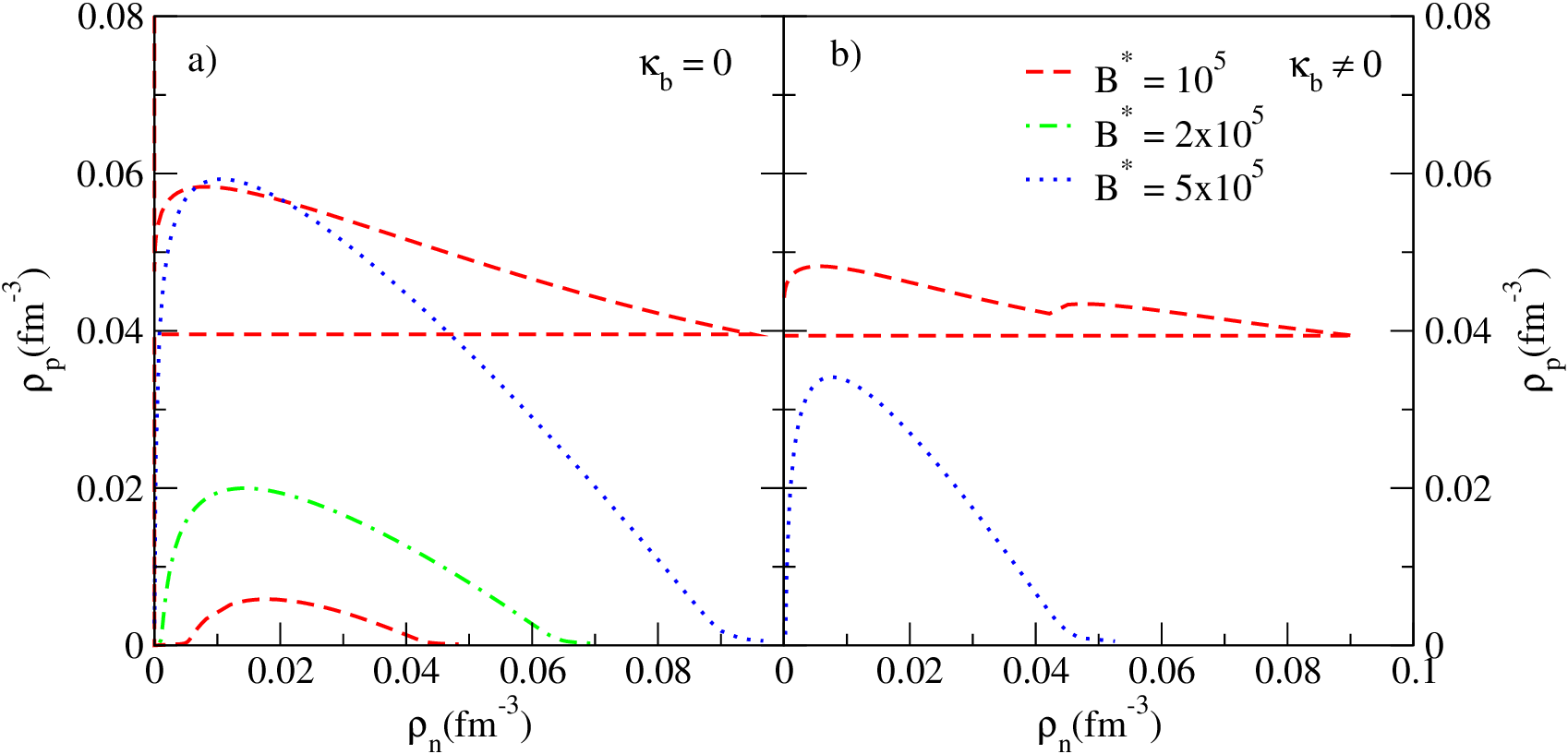}
\caption{(Color online) Spinodal section in terms of $\rho_p$ versus
  $\rho_n$ for TW for $npe$ neutral matter at $T = 0  \hbox{ MeV}$ and
  for severals values of magnetic fields (a) without  and  (b) with  AMM.}
\label{spin_npe_tw}
\end{figure}

For TW, contrary to the $B=0$ case,  the inclusion of the magnetic field gives
rise to a spinodal region as seen in  Fig. \ref{spin_npe_tw}.
The behavior of this model with the magnetic field  is similar to
TM1. We also point out that the
 inclusion of the AMM has a strong effect on the spinodal part corresponding to the first LL: it is drastically reduced or even disappears.

\subsection{Direction of instability}

\begin{figure}[htb]
\vspace{1.5cm}
\includegraphics[width=0.7\linewidth,angle=0]{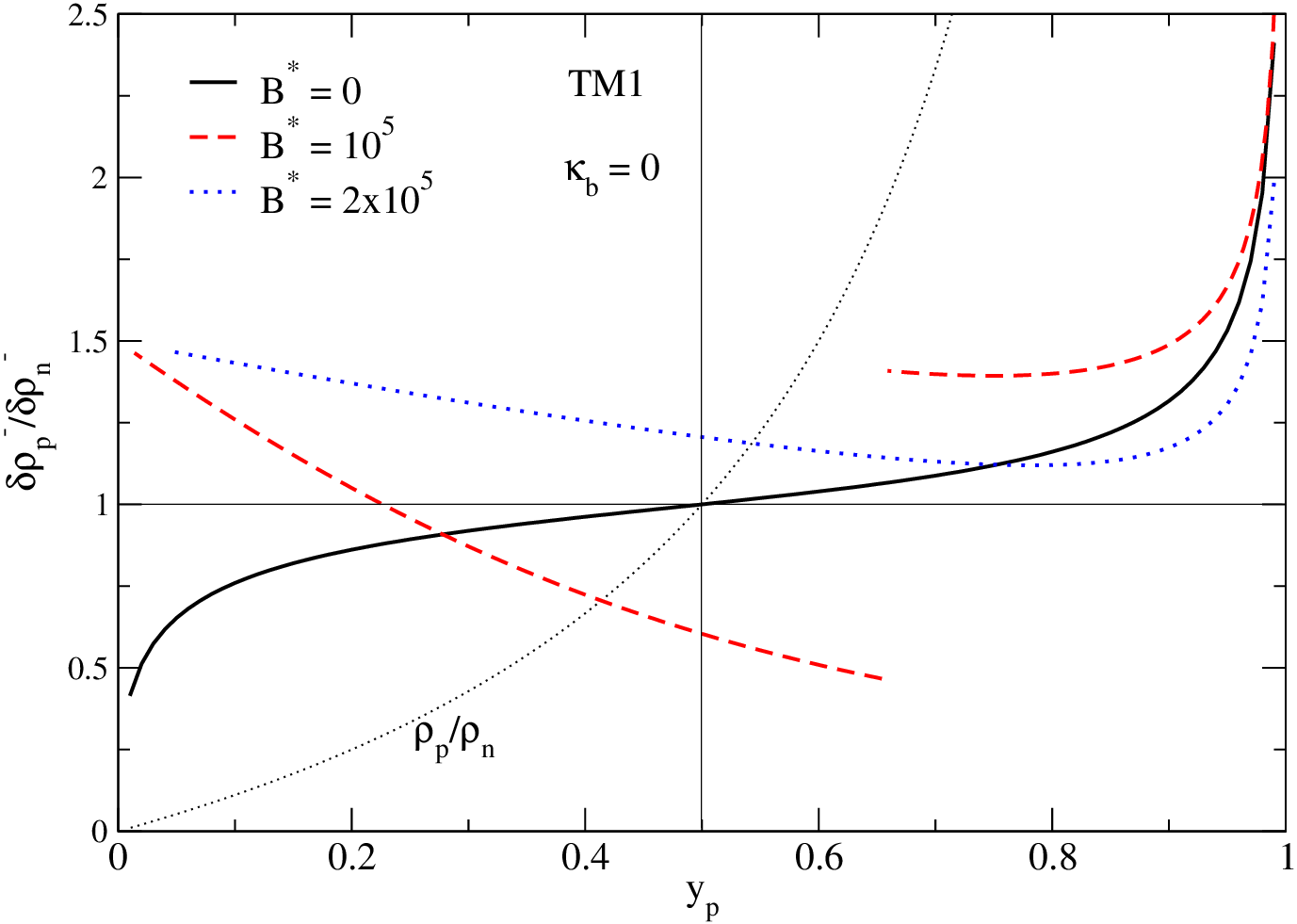}
\caption{(Color online) $\delta \rho^{-}_{p}/\delta \rho^{-}_{n}$ plotted as a
function of the proton fraction with $T = 0  \hbox{ MeV}$ and $\rho= 0.06
\hbox{ fm}^{-3}$  for the TM1 model and several values of the magnetic
fields  without  AMM. The fraction $\rho_p/\rho_n$ is given by the thin dotted line.}
\label{spdtm2}
\end{figure}

The eigenvector associated with the negative eigenvalue of the free energy curvature
matrix  defines the direction of the instability and tells us how does the system
separate into a dense liquid and a gas phase. It was shown in \cite{abmp06,camille08} that
in the absence of the magnetic field the direction of instability favors the
reduction of the isospin asymmetry of the dense clusters of the system, and
increases the isospin asymmetry of the gas surrounding the clusters, the so
called distillation effect. This effect is represented in Fig.~\ref{spdtm2}
where it is seen that for  the $B=0$ curve (thick full line)  the fraction $\delta
\rho^{-}_{p}/\delta \rho^{-}_{n}$
is larger than $\rho_p/\rho_n$ below $y_p=0.5$ and the other way round above.

In Fig. \ref{spdtm2} we plot, for TM1,  the ratio  $\delta \rho^{-}_{p}/\delta \rho^{-}_{n}$  for $\rho=0.06$ fm$^{-3}$ as a function of the proton fraction. Several results for different values of the magnetic field are shown by the thick lines. The thin lines represent the ratio $\rho_p/\rho_n$ for reference and $y_p=0.5$ points corresponding to symmetric matter, as well as   $\delta \rho^{-}_{p}/\delta \rho^{-}_{n}=1$, which is the ratio of density fluctuations for symmetric matter with no field. For the largest field considered the spinodal region contains a single
Landau level and the curve varies smoothly starting at  $\delta
\rho^{-}_{p}/\delta \rho^{-}_{n}\sim1.5$. We
point out the very large value of this fraction, always above 1, for
$y_p<0.5$.  The magnetic field favors an increase of the proton fraction
quite above the symmetric matter value. For $B^*=10^5$ the spinodal has two bands, see Fig.~\ref{spztm1npn} and \ref{spztwnp}
, corresponding to the occupation of the
first two Landau levels. The transition from one to the other is clearly seen
with  a large discontinuity of $\delta
\rho^{-}_{p}/\delta \rho^{-}_{n}$ at $y_p\sim 0.7$. Above this $y_p$ value the
curve behaves like the previous ones. However for $y_p<0.7$ the behavior is
quite different: the curve decreases from the value at $y_p$=0, which is
independent of the magnitude of the  magnetic field, to a value much smaller
than the corresponding value of the  fraction $\rho_p/\rho_n$. The fluctuations will not drive the system
out of the first Landau level and therefore the larger the proton fraction, the
closer the system comes to the top of the band and the smaller are the allowed
proton fluctuations. For $y_p>0.7$ or for the larger magnetic fields the
Landau levels are only partially filled and the fluctuations will never drive
the system out of the corresponding Landau level. 

\begin{figure}[htb]
\begin{tabular}{c}
\includegraphics[width=0.7\linewidth,angle=0]{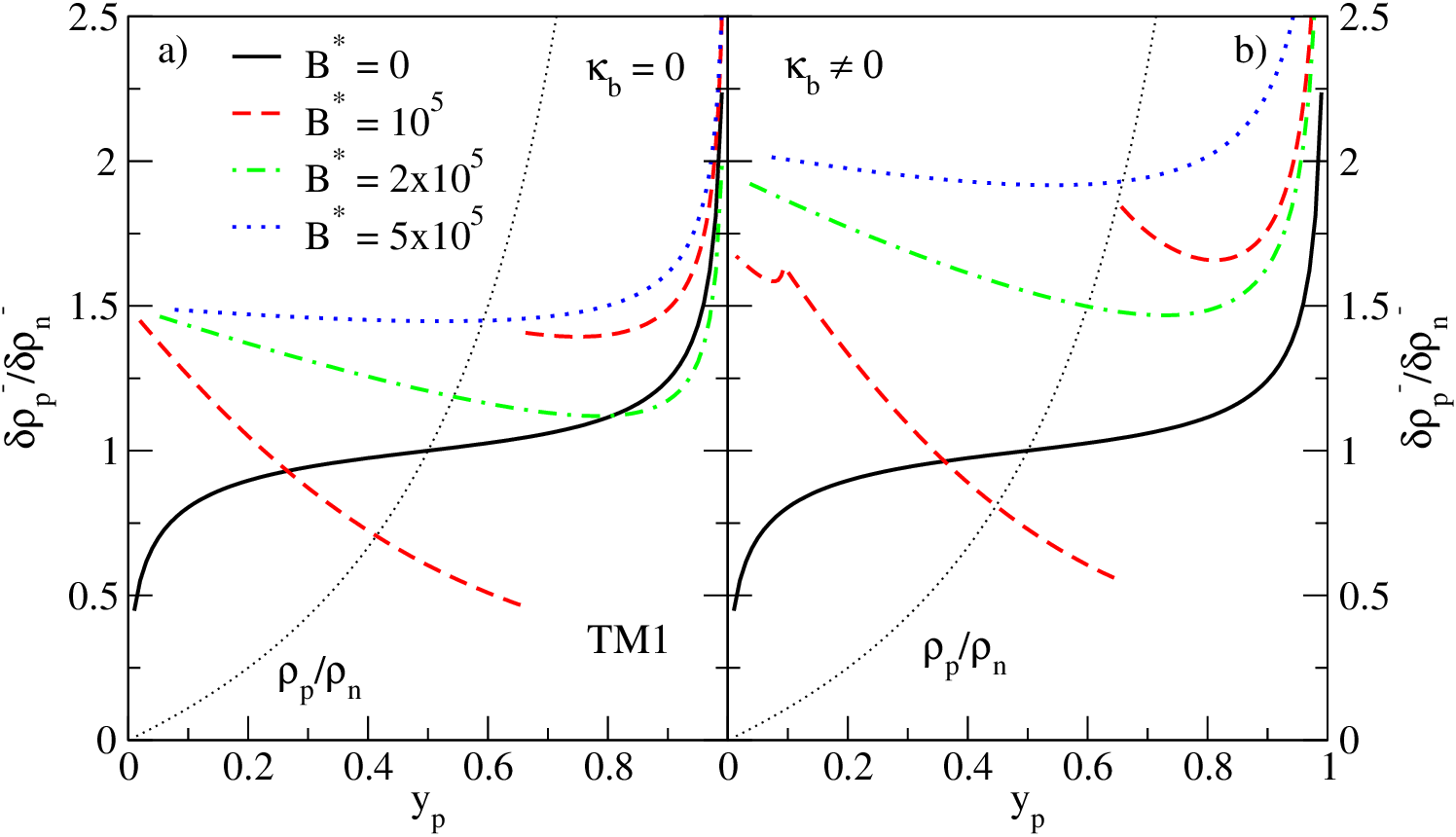}\vspace{1.5cm}\\
\includegraphics[width=0.7\linewidth,angle=0]{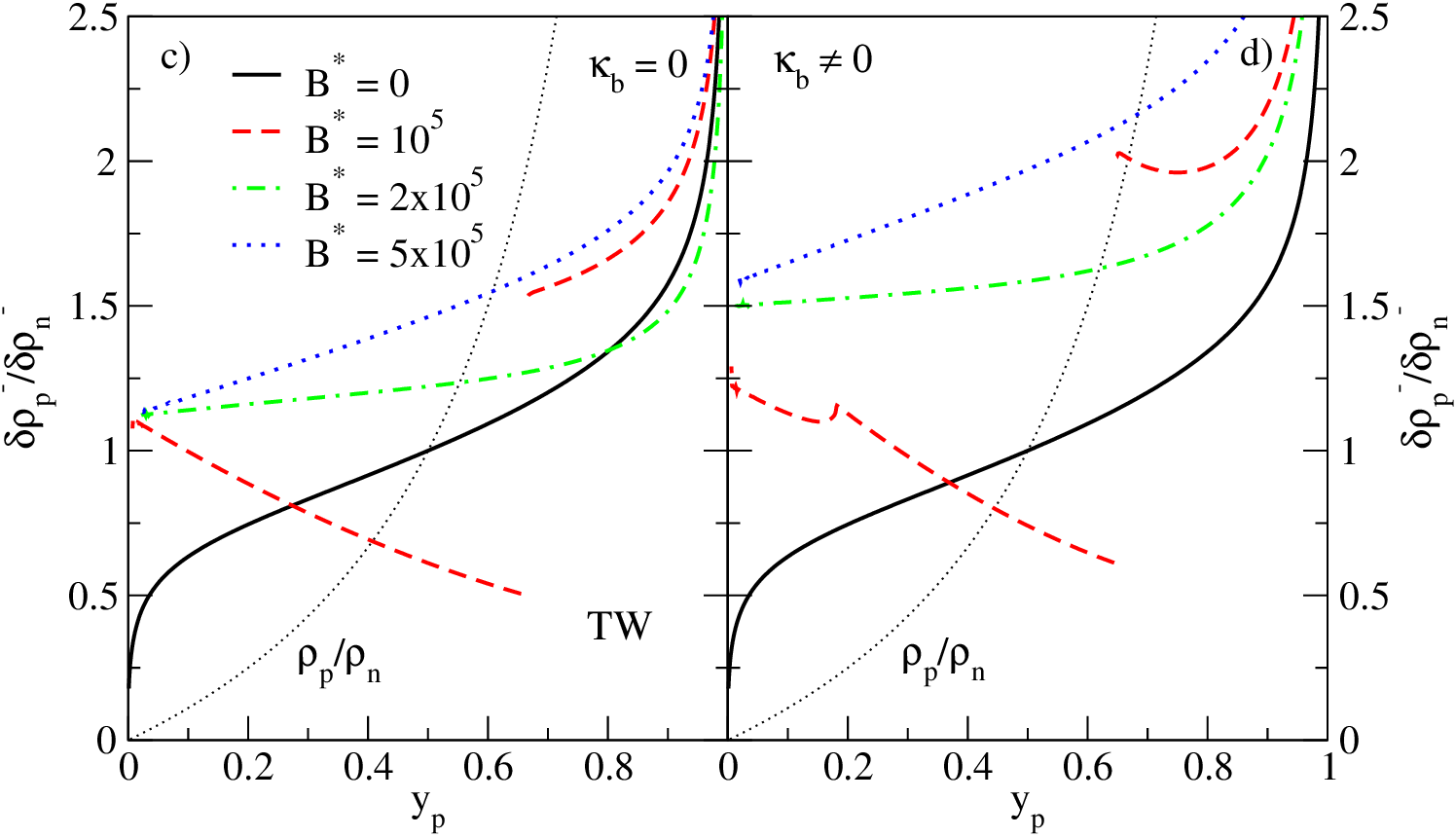}
\end{tabular}
\caption{(Color online) $\delta \rho^{-}_{p}/\delta \rho^{-}_{n}$ plotted as a
function of the proton fraction with $T = 0  \hbox{ MeV}$ and $\rho= 0.06
\hbox{ fm}^{-3}$  for the TM1 (top) and TW (bottom) and for severals values of the magnetic
fields  without (a) and (c) and with (b) and (d) AMM. The fraction $\rho_p/\rho_n$ is given by the thin dotted line.}
\label{spdtm1}
\end{figure}

Similar features are obtained for TW and/or including the AMM. In Fig. \ref{spdtm1}
  we show, respectively for TM1 (top)  and TW (bottom), the fraction $\delta
\rho^{-}_{p}/\delta \rho^{-}_{n}$ as a function of $y_p$ for a fixed baryonic
density, $\rho= 0.06 \hbox{ fm}^{-3}$, chosen inside the instability region.
For $y_p>0.5$, TM1 and TW behave in a similar way, while below this value the main 
difference is the smaller $\delta \rho^{-}_{p}/\delta \rho^{-}_{n}$ for TW, corresponding to a smaller distillation effect. This behavior is also present for $B=0$ and it was shown that this was due to the presence of the rearrangement term. The inclusion of the AMM favors larger proton fractions because neutron polarization stiffens the EOS.

\begin{figure}[htb]
\vspace{1.5cm}
\centering
\includegraphics[width=0.7\linewidth,angle=0]{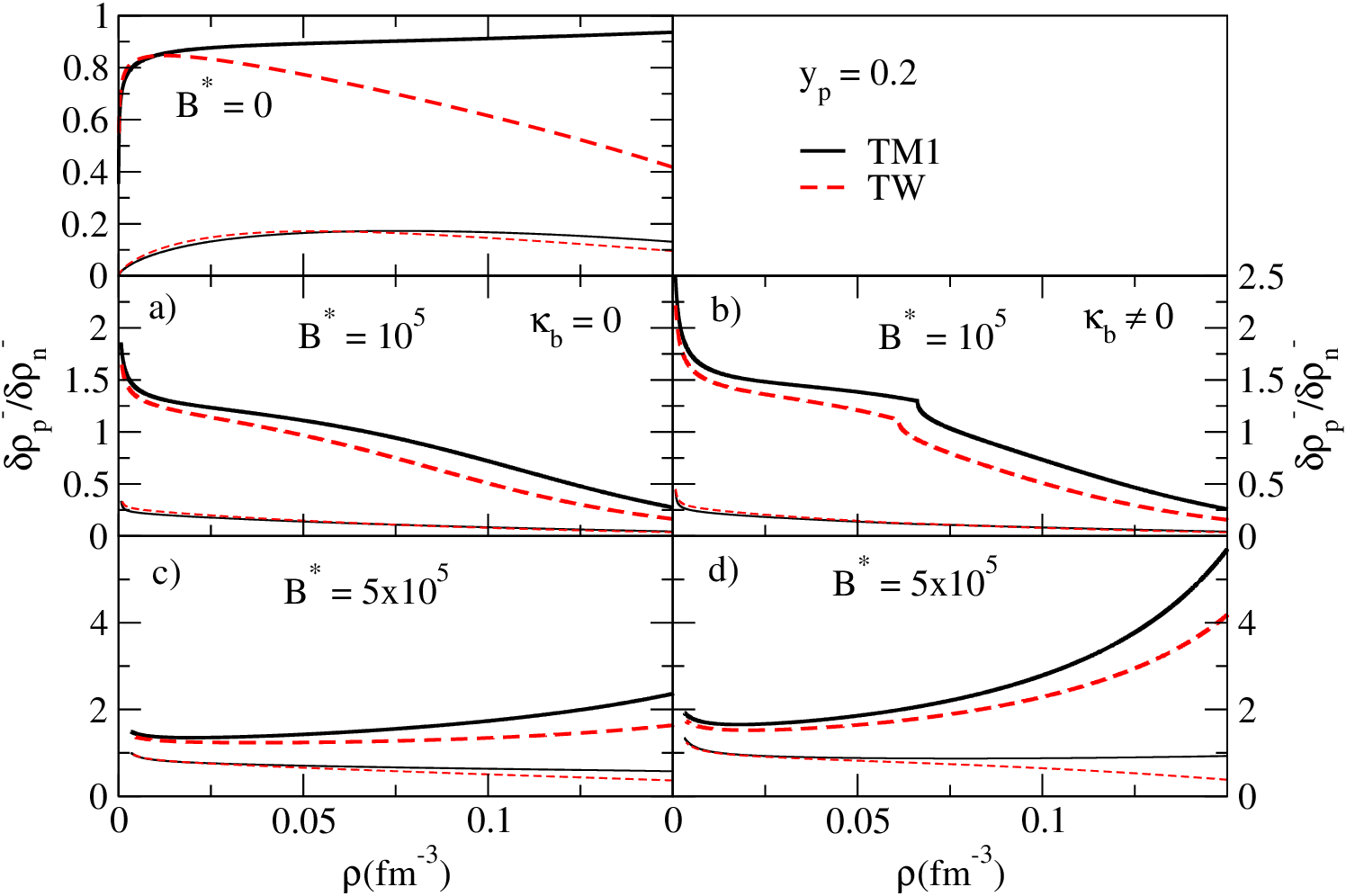}
\caption{(Color online) $\delta \rho^{-}_{p}/\delta \rho^{-}_{n}$
  plotted as a function of the density with $T = 0  \hbox{ MeV}$ and
  $y_p = 0.2$ with (thin lines) and without (thick lines) electrons
  for the TM1 and TW models and for severals values of the magnetic
  fields (a) and (c) without  and (b) and (d) with AMM.}
\label{drpdrnyp02}
\end{figure}

\begin{figure}[htb]
\centering
\includegraphics[width=0.7\linewidth,angle=0]{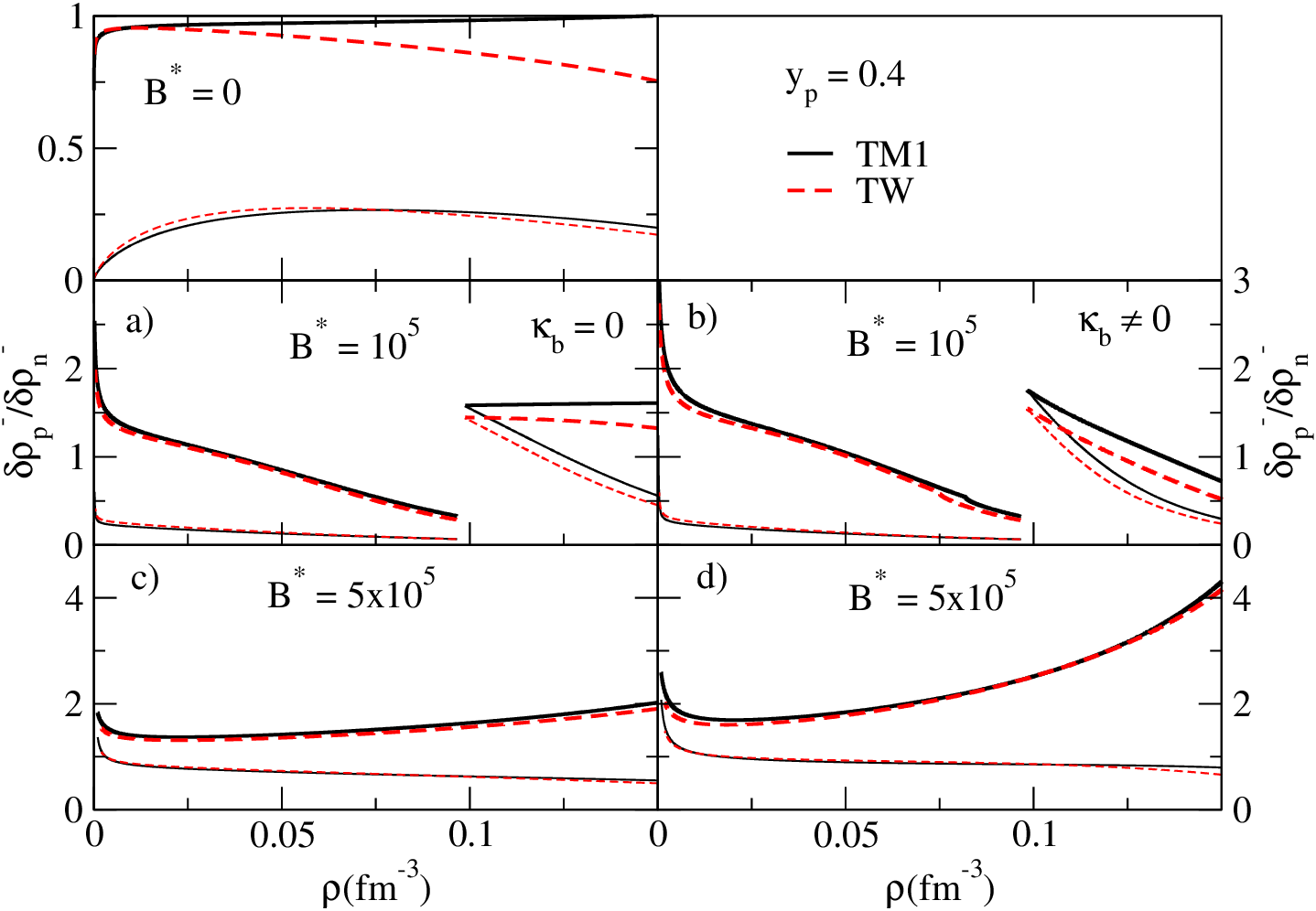}
\caption{(Color online) $\delta \rho^{-}_{p}/\delta \rho^{-}_{n}$
  plotted as a function of the density with $T = 0  \hbox{ MeV}$ and
  $y_p = 0.4$ with (thin lines) and without (thick lines) electrons
  for the TM1 and TW models and for severals values of the magnetic
  fields a)  without  and b)  with  AMM.}
\label{drpdrnyp04}
\end{figure}

In Figs.~\ref{drpdrnyp02} and~\ref{drpdrnyp04} we represent the fraction
$\delta \rho^{-}_{p}/\delta \rho^{-}_{n}$ as a function of density
respectively for two
values of $y_p$, 0.2 and 0.4, for $np$ matter (thick lines) and $npe$ matter
(thin lines). We consider TM1 and TW.
 Both models have
a very similar behavior for finite values of $B$ although for $y_p=0.2$ and $B=0$ they differ: for
TM1, $\delta \rho^{-}_{p}/\delta \rho^{-}_{n}$   increases with density while,
for TW, this fraction decreases for $\rho>0.02$ fm$^{-3}$. This effect is not
so strong for $y_p=0.4$ and for $npe$ matter the fraction is always quite small
due to  the presence of electrons which prevents large proton variations.

For $B=10^5$,  $\delta \rho^{-}_{p}/\delta \rho^{-}_{n}$ decreases with
density while for $B=5\times 10^5$ the opposite occurs. In both cases only the first
Landau level is occupied, however for the lower field the first Landau level is
almost full and the density
fluctuations will occur in such a way that the system stays in the same Landau
level: the larger the total density the smaller the fluctuations. For the
larger field the first Landau level is only partially filled, far the top of the band. For the same nuclear density, the larger the proton fraction
the lower the system energy and therefore the fraction $\delta
\rho^{-}_{p}/\delta \rho^{-}_{n}$ increases with density.

In Fig.~\ref{drpdrnyp04} we give the same information with $y_p=0.4$. While for
$B^* =5\times 10^5$ for the range of densities considered, matter occupies only
one Landau level, for $B^* =10^5$ we may have two (TM1) or three (TW) Landau
levels, see Figs.~\ref{spztm1npn} and~\ref{spztwnp}.  This explains the
discontinuities occurring for $\sim 0.1$ fm$^{-3}$. For the highest magnetic
field only one Landau level partially filled comes into play and
therefore the
fraction $\delta \rho^{-}_{p}/\delta \rho^{-}_{n}$ 
increases with density because that is favored energetically. For $B^* =10^5$
the presence of almost filled Landau levels prevents the existence of large
proton density fluctuations.

\section{Conclusions and outlook}

In the present work we have studied the instabilities of $np$ matter and $npe$ neutral matter under
very strong magnetic fields. The fields considered are much stronger than the
strongest field measured until now at the surface of a magnetar which is $B^*\sim 10^2$
for SGR 1806-20 \cite{sgr}. However, it is expected that  fields in the interior of neutrons
stars will be much larger. The present work shows how fields of the order
of $B=5\times 10^{18}$ G could affect the inner crust of a compact star.

We have considered  two relativistic nuclear models: one NLW model (TM1)  and one
DDRH model (TW). For both models, we have determined the spinodal surface, from the curvature
matrix of the free energy, for different magnitudes of the magnetic field. It
was shown that the instability region could be divided into several bands
according to the magnitude of the magnetic field and the number of  the Landau
levels occupied.  The presence of the magnetic field will generally  increase the
instability region. Making a crude estimation of the transition density at the
inner crust of a compact star under a strong magnetic field from the crossing
of the EOS of $\beta$-equilibrium stellar matter with the thermodynamical spinodal, it was shown that the transition
density and associated pressure increases with the magnetic field. This will
affect the structure of the star increasing the fraction of mass and of the
star' s moment of inertia  concentrated at the crust. These effects will be
noticeable if, for densities of the order of 0.1 fm$^{-3}$, the magnetic field
is  of the order of $B^*=10^4$ or larger.

The TW model has larger instability regions than  the TM1 model for the larger
proton densities. A smoother increase of the proton chemical potential for the
first model justifies this behavior. This behavior of the symmetry energy may  even give rise to a larger
number of bands in the spinodal of TW than the spinodal of TM1 for the same magnetic field.

We have also investigated the direction of instability. It was  shown
that if the Landau level is only partially occupied the density fluctuations
are such that the system evolves for a state with dense clusters very proton
rich immersed in a proton poor gas. A larger proton fraction is favored
energetically due to the degeneracy of the Landau levels. If on the other hand,
we study the fluctuation of particles occupying an almost complete Landau level, proton fluctuations
cannot be so large and it may even occur an anti-distillation effect with a
decrease of the proton fraction of the dense clusters. This is due to the fact
that these fluctuations will keep the system in the same Landau level.

\begin{acknowledgments}

This work was partially supported by FEDER and FCT (Portugal)
under the grant SFRH/BPD/14831/2003, and projects  PTDC/FP/64707/2006 and  POCI/FP/81923/2007. 

\end{acknowledgments}

\end{document}